\documentstyle[pre,epsf,aps]{revtex}

\begin{document}

\draft

\twocolumn[\hsize\textwidth\columnwidth\hsize\csname
@twocolumnfalse\endcsname
\title{Critical Dynamics of Self-Organizing Eulerian Walkers.}
\author{
R.R.Shcherbakov\cite{inst},
Vl.V.Papoyan,
A.M.Povolotsky}
\address{Bogoliubov Laboratory of Theoretical Physics,
Joint Institute for Nuclear Research,
141980 Dubna, Russia.}
\date{\today}
\maketitle
\begin{abstract}
The model of self-organizing Eulerian walkers is
numerically investigated on the square lattice.
The critical exponents for the distribution
of a number of steps ($\tau_l$) and visited sites ($\tau_s$)
characterizing the process of transformation from one
recurrent configuration to another
are calculated using the finite-size scaling analysis.
Two different kinds of dynamical rules are considered.
The results of simulations show that both the versions of the model
belong to the same class of universality with the critical exponents
$\tau_l=\tau_s=1.75\pm 0.1$.
\end{abstract}
]

To understand the nature and physical origins of the
Self-Organized Criticality (SOC)~\cite{BTW}, a number of different
models have been suggested in the last few years. Typically, these
models evolve according to the prescribed dynamical rules  into
the SOC state where they show spatio-temporal
self-similarity and are characterized by long range correlations.
In the critical state, the systems pass from one stable
configuration to another through the avalanches which play
an essential role in the organization of the dynamical
criticality.

In this letter, we numerically investigate
critical dynamics of a new cellular automaton model recently proposed
by Priezzhev, Dhar et al~\cite{PD}. This model has some common
features with the well-known Abelian Sandpile Model (ASM)~\cite{D,P}
but differs in rules which govern the motion of particles on
the lattice.

The model of the
self-organizing Eulerian walkers on the square lattice
is defined as follows. We associate with each site $i$ of the
two-dimensional $L\times L$ lattice an arrow
directed up, right, down or left from $i$. We start with an arbitrary
initial configuration of arrows on the lattice.
Initially, we drop a particle on the randomly chosen
site $i$. The succeeding steps the particle performs are determined
by the following rules:
\begin{enumerate}
\item[{\bf i)}] the particle coming to a site $j$ turns the arrow
                clockwise by the right angle,
\item[{\bf ii)}] then makes a step along the new direction of the
         arrow to the neighbor site,
\item[{\bf iii)}] if the new direction points out the lattice,
         the particle leaves the system.
\end{enumerate}

These rules are applied until the particle eventually leaves the lattice.
Then, we go on by adding a new particle and so on.
In this type of dynamics, the movement of the particle
affects the medium and in turn is affected by the medium.

On the lattice with closed boundary conditions,
the particle never leaves the system and
finally gets into a limit cycle in which it passes each
bond in both directions only once.
The walks of this type are known as Euler circuits~\cite{H}.

Let us consider the lattice with open boundary conditions.
The set of bonds marked by arrows form a graph $G$.
Adding the particles followed by their movement
through the lattice changes the configuration of arrows
and organizes the system into the SOC state~\cite{PD}.
This critical state is a collection of recurrent configurations
of a Markov process because the motion of a succeding particle
depends only on the final configuration of arrows produced
bu its predecessor.
Each element of this recurrent set can be obtained from
the previous one by adding a particle followed by
its movement and leaving the lattice.
It turns out that for any recurrent configuration, the graph $G$
is a spanning tree on the lattice. Thus, the set of recurrent
configurations is in the one-to-one correspondence with spanning
trees~\cite{PD}.

At each intermediate step, the moving particle can destroy a
spanning tree and form a loop of arrows. At this moment the
system leaves the recurrent set.
Eventually, after finite number of steps the particle reconstructs
the structure of a spanning tree.
The interval between the
destruction and restoration of the spanning tree can be called
an {\it avalanche of cyclicity}. During the walk of the particle,
the system passes several times from one recurrent configuration
to another through the avalanches of cyclicity.
This process is similar to the avalanche dynamics of sand in the
ASM where avalanches also reconstruct the recurrent configurations
(or spanning trees as the correspondence between the recurrent
set of ASM and spanning trees does exist as well).

At the beginning of an avalanche of cyclicity the last turned arrow
closes a loop. Then, the trajectory of the particle covers the
interior of the loop. During this walk each inner arrow turns four
times, whereas the arrows forming the loop turn in such
a way that the direction of the loop gets reversed.
It might occur that a moving particle, after escaping from one closed
loop, may form a new loop.
Due to this structure of the walk, the number of steps in an avalanche
is equal to $k(4n+1)$, where $k$ is the number of loops constituting
the avalanche and $n=0,1,\ldots$
It is possible to prove that the avalanche
may consist of only one or two loops. This fact explains
the line doubling in the distribution of the steps in the
avalanche of cyclicity (Fig.1).

To investigate the avalanche process in the models of
self-organizing Eulerian walkers, we studied them numerically
with high statistics. For each distribution
of avalanches we considered up to $30\cdot 10^6$ events on
the square lattices of linear size $L$ from $120$ to $400$.
Simulations always started from the regular initial
configuration in which all arrows were directed up.

In Fig.2, we present the double logarithmic plot
of the distribution $P(s)$ of the number of visited sites
in the avalanche for the lattice size $L=400$.
The analysis of the data shows that this distribution obeys
the power law
\begin{equation}
P(s)\sim s^{-\tau_s}\,.
\end{equation}

To estimate the critical exponents, we have performed the
finite-size scaling analysis~\cite{KNWZ,B} assuming
the distribution functions scale with the lattice size $L$ as
\begin{equation}
P(x,L)=L^{-\beta}f(x/L^{\nu})\,,
\end{equation}
where $f(y)$ is a universal scaling function and $\beta$ and $\nu$
are critical exponents which describe the scaling of the
distribution function.

To reduce the fluctuations of the data, we integrated each
distribution over exponentially increasing bin lengths.
For the integrated bin distribution we have~\cite{M}
\begin{equation}
D(s)=\int P(x)\,dx\sim s^{-(\tau_s-1)}\,.
\end{equation}

Plotting $D(s,L)L^{\beta_s}$ versus $s\,L^{-\nu_s}$ on a double
logarithmic scale, as is shown in Fig.3 for the different lattice
sizes $L$, we obtained that the best
data collapse corresponds to $\beta_s=1.5\pm 0.05$,
$\nu_s=2.0\pm 0.05$ (Fig.4).
The scaling relation for the critical exponents
$\tau_s={\beta_s}/{\nu_s}+1$ gives the value $\tau_s=1.75\pm 0.05$.

In the same way, we investigated the distribution
$P(l)$ of the number
of steps performed by the particle in the avalanche for the different
lattice sizes $L$.
There is an explicit power law behavior in these distributions (Fig.1)
\begin{equation}
P(l)\sim l^{-\tau_l}
\end{equation}
with the finite size cutoff.
We applied the finite-size scaling analysis to the integrated
distributions and obtained $\tau_l=1.7\pm 0.05$ from the best data
collapse.

We also investigated a slightly modified model. The difference from the
previous one is in the order of turns of the arrow.
In the case when the turns form the sequence (up-down-left-right-up),
we found a similar power law for avalanche distributions.

To find critical exponents of this power law from finite-size
scaling analysis, we integrated again these distributions over
exponentially increasing bin lengths.
The critical exponents $\tau_l=1.72\pm 0.05$ and $\tau_s=1.8\pm 0.05$
have been obtained from the best data collapse for the distribution
of steps and visited sites, respectively.

In conclusion, we numerically investigated the model of self-organizing
Eulerian walkers on the square lattice. The dynamics of the model organizes
the medium of the system and builds up spatio-temporal complexity.
We obtained explicit power law distributions in two slightly
different versions of the model.
We calculated the critical exponents for the distribution
of a number of visited sites $(\tau_s)$ and number of steps $(\tau_l)$
in avalanches of cyclicity.
These exponents are equal within a small uncertainty.
We argue that the critical exponents for these models within small errors
belongs to the same class of universality and
have a surprisingly large value $1.75\pm 0.1$ in comparison with the
known exponent for the ASM $(\tau=5/4)$~\cite{PKI}.

\vspace{2cm}

\begin{center}
ACKNOWLEDGMENTS
\end{center}

We would like to thank V.B.Priezzhev for valuable discussions and
critical reading of the manuscript.

One of us (R.R.S) was partially supported by
International Soros Science Educational Program.

\newpage

\onecolumn

\vspace*{40mm}

\begin{enumerate}
\item[FIG.1]
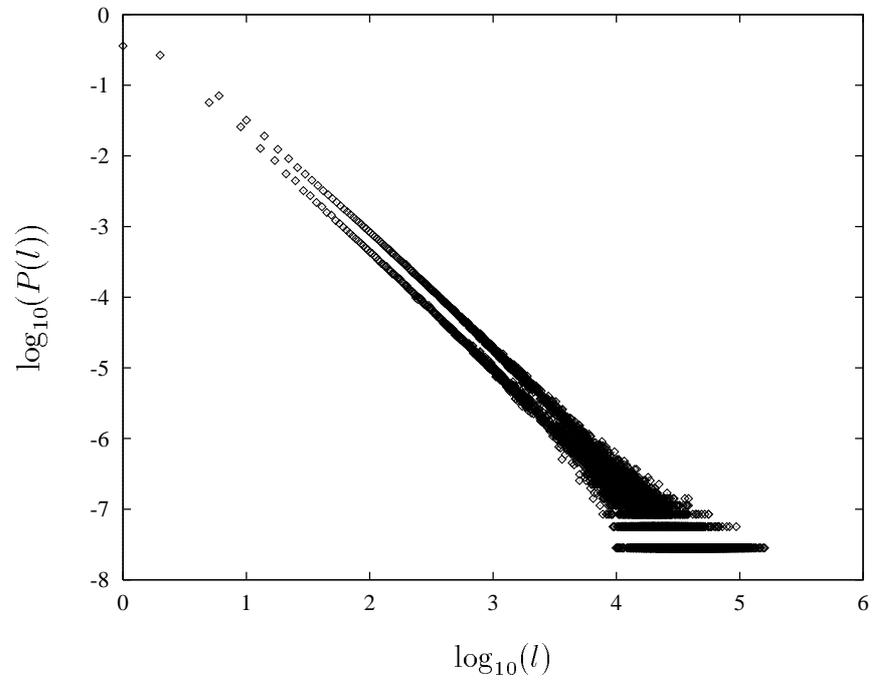
The distribution $P(l)$  of the number of steps
in avalanches on the square lattice of the linear size $L=400$.
\item[FIG.2]
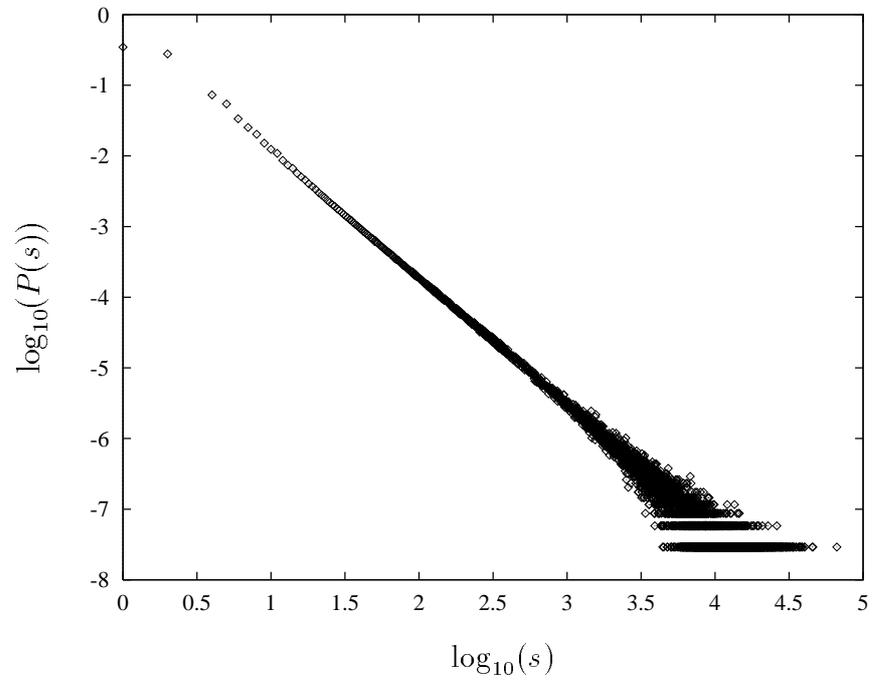
The distribution $P(s)$ of the number of visited sites
in avalanches on the square lattice of the linear size $L=400$.
\item[FIG.3]
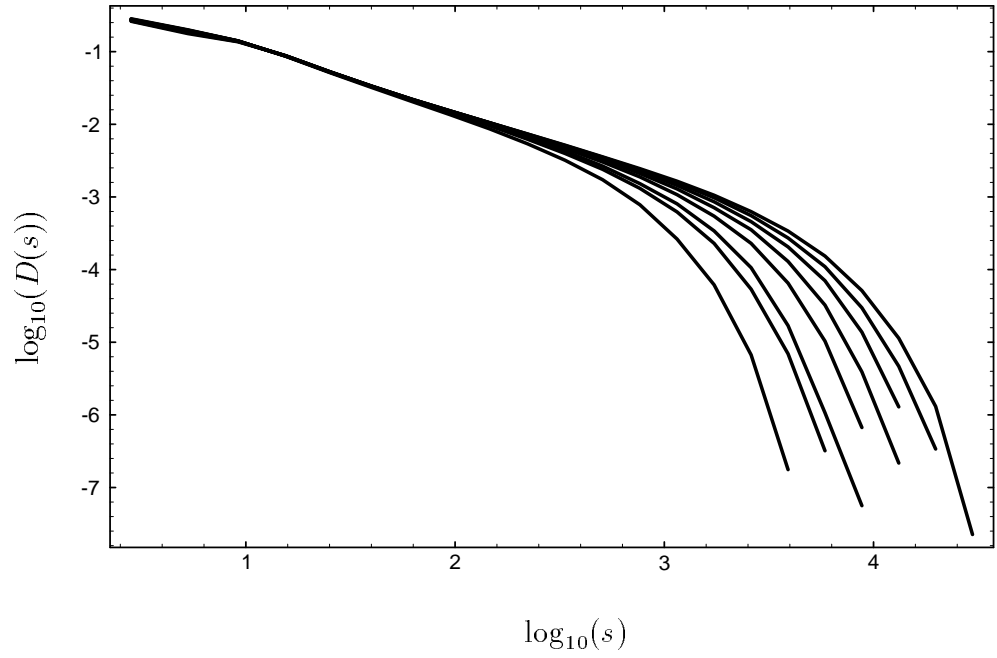
The integrated distributions $D(s)$ for the eight lattice
sizes with $L$ ranging from $120$ to $400$.
\item[FIG.4]
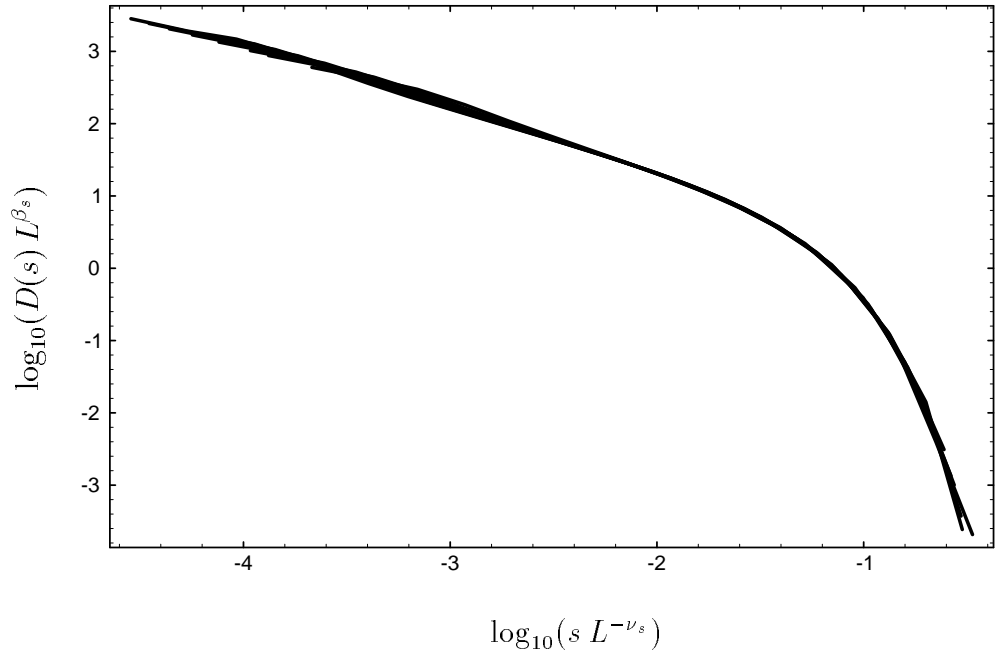
The finite-size scaling for the integrated distributions D(s).
\end{enumerate}


\begin{thebibliography}{99}
\bibitem[*]{inst}
On leave of absence from Theoretical Department, Yerevan Physics Institute,
375036 Yerevan, Armenia.
\bibitem{BTW}
                 P.Bak, C.Tang, and K.Wiesenfeld,
                 Phys. Rev. Lett. {\bf 59}, 381 (1987);
                 Phys. Rev. {\bf A 38}, 364 (1988).
\bibitem{PD}     V.B.Priezzhev, D.Dhar, A.Dhar and S.Krishnamurthy,
                 preprint TIFR, 1996.
\bibitem{D}
                 D.Dhar,
                 Phys. Rev. Lett. {\bf 64}, 1613 (1990).
\bibitem{P}
                 V.B.Priezzhev,
                 J. Stat. Phys. {\bf 74}, 955 (1994).
\bibitem{H}      F.Harary,
                 {\it Graph Theory}, MA: Addison-Wesley,
                 chapters 7 and 16, (1990).
\bibitem{KNWZ}   L.P.Kadanoff, L.Nagel, L.Wu, and S.Zhou,
		 Phys. Rev. {\bf A 39}, 6524 (1989).
\bibitem{B}      M.N.Barber,
                 in {\it Phase Transitions and Critical Phenomena},
                 edited by C.Domb and J.L.Lebowitz
                 (Academic, London, 1983), Vol.8, p.144.
\bibitem{M}      S.S.Manna,
                 Physica {\bf A 179}, 249 (1991).
\bibitem{PKI}
                 V.B.Priezzhev, D.V.Ktitarev, E.V.Ivashkevich,
                 Phys. Rev. Lett. {\bf 76}, 2093 (1996).
\end{thebibliography}
\end{document}